\documentstyle[12pt]{article}

\newcommand{\la}[1]{\label{#1}}

\newlength{\numlen}

\newlength{\indexlength}

\newcommand{\be}{\begin{equation}}
\newcommand{\ee}{\end{equation}}
\newcommand{\ba}{\begin{eqnarray}}
\newcommand{\ea}{\end{eqnarray}}
\newcommand{\bb}{\begin{array}}
\newcommand{\eb}{\end{array}}

\newcommand{\eq}[1]{Eq.~(\ref{#1})}
\newcommand{\eqs}{Eqs.~}

\begin{document}

\begin{titlepage}
\hfill BI-TP-93/31\\
\mbox{}\hfill AZPH-TH/93-19\\
\mbox{}\hfill SPhT 93/053
\begin{centering}
\vfill

{\bf CRITICAL BEHAVIOUR OF THE 3D GROSS-NEVEU}

{\bf AND HIGGS-YUKAWA MODELS}

\vspace{1cm}
L. K\"arkk\"ainen$^a$, R. Lacaze$^b$, P. Lacock$^c$, and B. Petersson$^c$

\vspace{0.3cm}
{\em $^a$Department of Physics, University of Arizona,\\
Tucson, AZ 85721, USA\\}
\vspace{0.3cm}

{\em $^b$Service de Physique Th\'eorique de Saclay,\\
91191 Gif-sur-Yvette Cedex, France.\\}

\vspace{0.3cm}
{\em $^c$Fakult\"at f\"ur Physik, Universit\"at Bielefeld,\\
Postfach 100131, 33501 Bielefeld, Germany\\}
\vspace{0.3cm}

\vspace{2cm}
{\bf Abstract}

\end{centering}

\vspace{0.3cm}\noindent
We measure the critical exponents of the three dimensional
Gross-Neveu model with two four-component fermions.
The exponents are
inferred from the scaling behaviour of observables on
lattice sizes $8^3$, $12^3$, $16^3$, $24^3$,
and $32^3$.  We find that the model has
a second order phase transition with $\nu = 1.00(4)$ and
$2-\eta = \gamma / \nu = 1.246(8)$. We also calculate
these exponents, through a second order $\epsilon$-expansion around four
dimensions, for the three dimensional Higgs-Yukawa model, which is
expected to be in the same universality
class, and obtain
$\gamma / \nu = 1.237$ and $\nu = 0.948$, while
recent second order $1/N_f$-expansion calculations
give $\gamma/\nu = 1.256$ and $\nu = 0.903$.
We conclude that the equivalence of the two models remains valid
in 3 dimensions at fixed small $N_f$ values.

\vfill \vfill
\noindent
BI-TP-93/31\\
AZPH-TH/93-19\\
SPhT 93/053\\
July 1993\\
\end{titlepage}

\section{Introduction}
The Gross-Neveu (GN) model describes fermions
with four-fermion interaction \cite{Gross74,Wilson73}.
It has a global discrete chiral symmetry,
which can break down spontaneously to form a chiral condensate.
This can be seen as a composite scalar particle that gives a
non-zero mass for the fermions.
Due to its simplicity, the GN model has been studied extensively.
In two dimensions it is perturbatively renormalisable
and asymptotically
free. In addition, the chiral symmetry is broken.

In three dimensions (3d) it is renormalisable if the number of
flavours, $N_f$, is large enough.
Hence, it is also the first renormalisable model
known not to be perturbatively renormalisable. It has been proven
to be renormalisable and non-trivial in dimensions
between 2 and 4 by means of a
$1/N_f$-expansion \cite{Wilson73}.

It has been suggested that a model with four-fermion interactions,
which leads to a spontaneously broken global chiral symmetry
with a chiral condensate, could be a candidate for the Higgs particle
of the standard model \cite{Bardeen86}.
Near four dimensions (4d) it has been related to
Higgs-Yukawa (HY) type models \cite{Hasenfratz91,Justin92}.
In fact, in 4d
both the standard electroweak and the GN model
(with certain modifications) are trivial
and can be mapped onto each other \cite{Hasenfratz91} and in
dimensions $2\le {\rm d}\le 4$ the GN model and the HY
model are equivalent in the framework of
$1/N_f$-expansion \cite{Justin92}.

The purpose of this work is to enlighten the
connection between the composite and fundamental
Higgs scenarios in a case where the models are not trivial,
namely in 3d.
The $1/N_f$ expansions for the HY model and the GN model
are order by order equivalent \cite{Justin92}. In order to go
beyond the $1/N_f$ perturbation theory, we ask whether this
equivalence persists at small
$N_f$'s. To answer this we study the critical properties of GN model
at small $N_f$ by means of a Monte Carlo (MC) simulation.
To preserve discrete chiral symmetry,
the restoration of which we are interested in,
we use staggered fermions on the lattice and choose
the smallest possible value of $N_f$, that is $N_f=2$. A finite-size
analysis of the numerical simulation gives the critical
coupling and critical exponent values to be compared to the
exponents of the Higgs-Yukawa model.
In order to obtain a meaningful comparison, one has to
go beyond the order $\epsilon$ results in the HY model
\cite{Justin92}. We have therefore extended the calculation
and obtain the critical exponents to
order $\epsilon^2$. The relevance of the comparison
can furthermore be evaluated by using recent results
from a $1/N_f$-expansion of
the GN model \cite{Hands92,Gracey91,Gracey92,sacha,Gracey93}
to order $1/N_f^2$.

The paper is divided into two major parts: section 2 is devoted
to the analytical results of the Gross-Neveu and Higgs-Yukawa models
with the $\epsilon$-expansion
while in section 3 we describe the Monte Carlo
runs performed. Section 4 is devoted to the comparison of numerical and
analytical results. Within the uncertainties associated on the one hand
with the statistics of the numerical simulations and
on the other hand with the still short series expansion
in $\epsilon$ and $1/N_f$, we confirm the equivalence of the two
3d models at small $N_f$.

\section{Analytical Results}
Here we recall a few known properties of the models under study, and
present our new analytical calculations, in particular the fixed $N_f$
$\epsilon^2$ expansion of the HY model exponents. The $1/N_f$ expansions
of the latter are then compared with the $1/N_f$ expansions of the
GN model.
\subsection{Continuum Gross-Neveu model}

The continuum GN model with $N_f$ fermion flavours is
defined by the Lagrangian
\be
{\cal{L}}=\sum_{\alpha =1}^{{N}_{f}}
{\overline{\psi }}^{\alpha }(x)\not{\hskip -2pt \partial}
{\psi }^{\alpha }(x)+{g^{2}\over
2}{\left[{\sum_{\alpha =1}^{{N}_{f}}  {\overline{\psi}}^{\alpha}
(x){\psi }^{\alpha }(x)}\right]}^{2} .
\la{eq1}
\ee

Usually an auxiliary scalar field $\sigma$ is introduced
\be
{\cal{L}}=\sum_{\alpha =1}^{{N}_{f}}{\overline{\psi }}^{\alpha }(x)
[\not{\hskip -2pt \partial}-g\sigma(x)]
{\psi }^{ \alpha }(x)-{1 \over 2}{ \sigma }^2(x) ,
\la{eq1a}
\ee
which is formally equivalent to (\ref{eq1}) upon integration over
$\sigma$ field.

The GN model has a discrete chiral symmetry
\be
\psi\rightarrow\gamma_5 \psi, \quad\overline{\psi }\rightarrow-
\overline{\psi }\rm \gamma_5, \quad\sigma\rightarrow-\sigma ,
\la{1d}
\ee
which in 3d is spontaneously broken at small couplings.

The critical exponents for $d=3$ to order
$1/N_f^2$ are \cite{Gracey92,sacha,Gracey93}
\be
{1\over\nu}  =  1-{32\over3 \pi^2 N}+{64(27\pi^2+632)\over27\pi^4N^2},
\la{nun}
\ee
\be
{\gamma\over\nu}  =  1+{64\over3 \pi^2 N}+{64(27\pi^2-304)\over27\pi^4N^2},
\la{gamn}
\ee
where $N=N_f \ {\rm Tr} \ {\bf 1}$ is
the total number of fermionic variables.

\subsection{The discretised Gross-Neveu model}
We consider the GN model defined on a 3d symmetric lattice.
The discretisation of the continuum Lagrangian has to
be implemented in such a way as to reproduce the correct
symmetries in the  continuum limit \cite{thierry}.
We use staggered fermions, which means that in 3d there
are 8 doublers. Using 4 component spinors, we can assign the components
to the corners of the cube, leaving $8/4 = 2$ continuum flavours
from 1 staggered lattice fermion per site. The discretised action reads
\be
S\rm =N_L \sum_n{{\sigma }_n^2 \over 2 \lambda }
+\sum_{n,m}\sum_{\alpha =1}^{{N}_{L}}
\overline{\chi }_n^{\alpha}({D}_{nm}+{\Sigma }_{nm}){\chi }_m^{\alpha}.
\ee
The staggered fermion matrix $D$ is given by
\be
{D}_{nm}={\frac{1}{2}}
\sum_{j} {\eta_{n,j}({\delta }_{n,m+\hat{j}} - {\delta }_{n,m-\hat{j}}}),
\ee
where the sum $j$ is over the directions ($j$ = 1,2,3) and
$\eta_{n,j}$ is the staggered fermion phase factor
\be
\eta_{n,j}= (-1)^{n_1 + ... + n_{j-1}} .
\ee
with periodic boundary condition in d-1 dimensions and antiperiodic
in the last one for a finite lattice.

\noindent The mass matrix $\Sigma$ is diagonal
\be
{\Sigma }_{nm} = {\bar \sigma} \delta_{nm}
\ee
and depends on $\sigma$ field.
We choose a discretisation in which
$\bar \sigma$ is the average of $\sigma$ at the
six nearest neighbours of the lattice site $n$.

\noindent The coupling $\lambda$ is connected to the continuum coupling
$g$ by $\lambda = g^2 N_f$, with $N_f=2 N_L$ as explained.

One may integrate over the Grassmann
variables $\chi^{\alpha}_x, \bar \chi^{\alpha}_x$ to express the partition
function in terms of an effective action
\be
S_{eff} = N_L \  \bigl[\sum_x {{\sigma^2_x} \over {2 \lambda}} - {\rm Tr}~\ln
(D + \Sigma) \ \bigr].
\la{seff}
\ee
The $N_L=\infty$ critical value $\lambda^0_c$, where
chiral symmetry is restored,
can be obtained from the saddle point equation at $\sigma=0$ \cite{Hands92} as
\be
\lambda^0_c=0.989.
\la{l0}
\ee
Taking into account the quadratic fluctuations \cite{belanger} we can
obtain the one loop value of $\lambda_c$ including $1/N_L$ corrections.
As we are interested to rather small $N_L$ values, we choose to solve
the gap equation of the one-loop effective potential. Solving this
gap equation on the lattice with
linear sizes $L=32 \ {\rm and} \ 64$ we estimate
\be
\lambda^1_c(N_L=1)=0.800
\la{l11}
\ee
\be
\lambda^1_c(N_L=6)=0.938
\la{l16}
\ee
This last value agrees with a direct
calculation of the perturbative correction in the infinite lattice limit.
\subsection{The Higgs-Yukawa model}
The Lagrangian ${\cal{L}'}$ of the HY
model has a fourth order
interaction term and a kinetic term added for the $\sigma$ field,
\be
{\cal{L}'}= \sum_{\alpha =1}^{{N}_{f}}
\overline{\psi }^\alpha(\not{\hskip -2pt \partial}
 + g \sigma) \psi^\alpha +{1 \over 2}
m^2 \sigma^2+{1 \over 2}(\partial_{\mu}\sigma)^2 +
{\frac{\lambda_4}{4!}}\sigma^4 .
\la{lhy}
\ee
This model becomes
renormalisable in four dimensions. It
has been argued that the terms added are
irrelevant for the number of dimensions less than 4 and marginal
at d=4. Hence, the critical behaviour of HY model and
GN model should be identical.

Due to its renormalisability, the HY model can be studied by means of
an $\epsilon$-expansion. The two-loop $\beta$ functions and anomalous dimension
$\eta_{\sigma}$ can be obtained from \cite{machacek} and we have computed the
mass anomalous dimension $\eta_m$.
\be
\beta_{\lambda_4}=-\epsilon \lambda_4
  +{1\over(4 \pi)^2} \bigl( 3 \lambda_4^2+2 N \lambda_4 g^2-12 N g^4 \bigr)
  +{1\over(4 \pi)^4}
 \bigl({-17\over3}\lambda_4^3-3 N \lambda_4^2 g^2+7 N
\lambda_4 g^4 +96 N g^6 \bigr) ,
\ee
\be
\beta_g=-{\epsilon \over 2} g+{1\over(4 \pi)^2} (N/2 +3) g^3
  +{1\over(4 \pi)^4} \bigl(-{9+12 N \over 4} g^5 -2
\lambda_4 g^3 +{1\over 12} \lambda_4 ^2 g\bigr) ,
\ee
\be
\eta_{\sigma}={1\over(4 \pi)^2} N g^2 +
     {1\over(4 \pi)^4}\bigl({1\over6} \lambda_4^2-{5\over2} N g^4\bigr),
\ee
\be
\eta_m=-{1\over(4 \pi)^2} \lambda_4
   +{1\over(4 \pi)^4}
 \bigl(\lambda_4^2+\lambda_4 N g^2-2 N g^4\bigr) -\eta_{\sigma},
\ee
\noindent with $d=4-\epsilon$ and $N=N_f \ {\rm Tr} \ {\bf 1}$.

The corresponding fix points to order $\epsilon^2$ are
\be
{g^{*2}\over(4 \pi)^2}= {1\over (N+6)} \epsilon
   +{(N+66) \sqrt{N^2+132 N+36}-N^2+516 N+882
  \over 108 (N +6 ) ^3} \epsilon^2
\ee
\ba
{\lambda_4^*\over(4 \pi)^2} = & &
 {-N+6+\sqrt{N^2+132 N+36}\over 6(N+6)}\epsilon\nonumber\\
 {}& +  &
  \bigl[\quad-\sqrt{N^2+132 N+36} \ (3N^3-43 N^2-1545 N-1224) \nonumber\\
 {}&  &\quad
  {\quad +3N^4+155N^3+2745 N^2-2538 N+7344\bigr] \over
    54 (N+6)^3 \sqrt{N^2+132 N+36}} \epsilon^2.
\label{lstar}
\ea
The anomalous dimensions at these fix points give the critical exponents
$1/\nu=2+\eta_m(g^*,\lambda_4^*)$
and $\gamma/\nu=2-\eta_{\sigma}(g^*,\lambda_4^*)$
\ba
{1\over\nu}=2&-&{5 N +6+\sqrt{N^2+132 N+36}\over 6 (N+6)}\epsilon\nonumber\\
  {} & - &
    \bigl[\quad\sqrt{N^2+132 N+36}(3N^3 + 109 N^2 + 510 N + 684) \nonumber\\
 {}&  &\quad
  {\qquad - 3 N^4 - 658 N^3 - 333 N^2 -15174 N + 4104\bigr] \over
    54 (N+6)^3 \sqrt{N^2+132 N+36} }\epsilon^2,
\label{nue}
\ea
\be
{\gamma\over\nu}= 2-{N \over N +6}\epsilon
   -{(11 N+6) \sqrt{N^2+132 N+36}+52 N^2-57 N+36\over 18 (N+6)^3} \epsilon^2.
\label{game}
\ee
\subsection{Gross-Neveu and Higgs-Yukawa Comparison}

 The $\epsilon=4-d$ expansions of the Gross-Neveu exponents
\cite{Gracey92,sacha,Gracey93} are
\ba
{1\over\nu}|_{GN}=2&-&\epsilon+
  (-6\epsilon+{13\over2}\epsilon^2-{3\over8}\epsilon^3+ \cdots )
{1\over N} \nonumber\\
 {} & + & \quad
 (396\epsilon-{1125\over2}\epsilon^2-{1140\zeta(3)-401\over8}
\epsilon^3+ \cdots ) {1 \over N^2},
\label{nune}
\ea
\ba
{\gamma\over\nu}|_{GN}  =  2 &-&\epsilon +
     ( 6\epsilon-{7\over2}\epsilon^2-{11\over8}
\epsilon^3+ \cdots){1\over N} \nonumber\\
    & + & (-36\epsilon+{51\over2}\epsilon^2+
{192\zeta(3)+281\over8}\epsilon^3+ \cdots){1\over N^2},
\label{gamne}
\ea
while the $1/N$ expansion of \eqs(\ref{nue} and \ref{game}) gives
\ba
{1\over\nu}|_{HY}=2&-&\epsilon+
  (-{6\over N}+{396\over N^2}-{26136\over N^3}+ \cdots)\epsilon \nonumber\\
 {} & + & \quad
 ( {13\over 2 N}-{1125\over 2 N^2}+{48951\over  N^3}+ \cdots)\epsilon^2,
\label{nuen}
\ea
\ba
{\gamma\over\nu}|_{HY}=2&-&\epsilon+        
  ( {6\over N}-{36\over N^2}+{216\over N^3}+ \cdots)\epsilon \nonumber\\
 {} & + & \quad
 ( -{7\over 2 N}+{51\over 2 N^2}+{1215\over N^3}+ \cdots)\epsilon^2.
\label{gamen}
\ea
Up to order $\epsilon^2$ and $1/N^2$ the two models agree as expected.
In the GN $1/N$-expansion, the $\epsilon^2$ terms are comparable to
the $\epsilon$ ones and the $\epsilon^3$ is relatively small in $1/\nu$
and of the same magnitude in $\gamma/\nu$.
In contrast, the HY $\epsilon$-expansion shows that the coefficients
of the $1/N$ expansion are always rapidly increasing, in particular
in the case of $1/\nu$. Thus a
resummation for the GN $\nu$ and the HY $\gamma/\nu$ has to be made to
improve the corresponding estimates. The necessity of such a resummation
is also manifest from the importance, at low $N$, of the $1/N_f^2$
contribution for the GN $\nu$, and
the $\epsilon^2$ one for the HY $\gamma/\nu$
(about 20\% for $N$=8).

Because of the lack of
information on asymptotic behaviour,
we use a simple Pad\'e-Borel resummation \cite{Hikami}
with arbitrary choice of function,
instead of the more
sophisticated Borel resummation \cite{Leguil}.
For an expansion
\be
A(x)=1+a_1 x +a_2 x^2 +{\cal O} (x^3),
\ee
we  write
\be
A(x)={1\over x} \int_0^{\infty}{\rm d}t {\rm e}^{-t/x}
  \bigl[1-a_1 t - (a_2/2-a_1^2) t^2 \bigr]^{-1}.
\label{borel}
\ee
These formulae are directly used for the
GN $\nu$ obtained from \eq{nun} \cite{Gracey93},
while we first expand the HY $\gamma/\nu$ of \eq{game} around
the $N=\infty$ point as
\be
{\gamma\over\nu}|_{HY}=
(2-\epsilon)(1+a_1\epsilon+a_2\epsilon^2)+{\cal O}(\epsilon^3),
\ee
and resum with \eq{borel} only the second bracket.

The comparison of the resulting critical
exponents for the two 3d models as a function
of the fermion number $N$ is summarised in Fig. 1, where the
dotted lines represent the
computation of the GN $1/\nu$ and the HY $\gamma/\nu$
without the resummation procedure described above.
The difference between the two models is small
except for $\nu$ at low $N$.
The data point
at $N=48$ comes from Ref. \cite{Hands92}, while those at $N=8$
result from the simulation described in the next section.

\section{Numercial Results}
Here we present our simulation and the analysis leading to estimates
of the critical indices for the 3d Gross-Neveu model at $N=8$.
\subsection{Simulation of the lattice Gross-Neveu model}

For the numerical simulation we consider the effective action \eq {seff} with
$N_L=1$ which corresponds to $N=8$, and use an
exact Hybrid Monte Carlo algorithm.
It has a point update of 8$\mu$s-14$\mu$s on a Cray Y-MP,
increasing with lattice size.
This is due to the fact that more conjugate gradient steps are
needed to invert the fermion matrix for large lattices.
We perform runs on lattice sizes $8^3$, $12^3$,
$16^3$, $24^3$ and $32^3$.  We use
$20$ time steps of length 0.2, except for the
largest lattice where the
time step is reduced to 0.05.
As a rule, measurements are carried out every 5th trajectory.
Details regarding the runs are
listed in Table~ \ref{table1}.
The integrated autocorrelation time $\tau_{int}$ quoted is that for
$ \left\langle{\sigma^2}
\right\rangle $.
\begin{table} [h]
\center
\caption[1]{Statistics of the simulations.}
\label{table1}
\begin{tabular}{lrrr}
\hline
Size&$ \lambda$&Trajectories&$\tau_{int}$\\
\hline
$8^3$ & $0.7875000$ & $110 000$ & $8$\\
$8^3$ & $0.8156250$ & $110 000$ & $11$\\
$8^3$ & $0.8437500$ & $190 000$ & $18$\\
$12^3$& $0.7875000$ & $ 90 000$ & $8$\\
$12^3$& $0.8156250$ & $430 000$ & $11$\\
$12^3$& $0.8437500$ & $ 90 000$ & $15$\\
$16^3$& $0.7875000$ & $ 90 000$ & $9$\\
$16^3$& $0.8156250$ & $282 000$ & $11$\\
$16^3$& $0.8437500$ & $100 000$ & $17$\\
$24^3$& $0.8156250$ & $432 240$ & $14$\\
$24^3$& $0.8184375$& $190 000$ & $4$\\
$32^3$& $0.8167500$& $267 160$ & $5$ \\
\hline
\end{tabular}
\end{table}

To analyse the data we use a variant of multihistogram
reweighting analysis
which does not require the binning of data \cite{kajantie91}.
This is used to obtain the values of observables in
between the simulated data points. These points
are close enough, and the simulations long enough, to produce
bosonic energy distributions that fill the whole coupling range
of interest.

\subsection{The scaling and critical exponents from MC data}

Let us first consider the critical coupling and the
critical exponent $\nu$, which describes the
behaviour of the correlation length near the phase transition.
We define the renormalised coupling $g_R$ as
\be
g_R \equiv {{\left\langle{\sigma^2}
\right\rangle}^2 \over \left\langle{\sigma^4}\right\rangle}.
\ee
This expression for $g_R$ lacks a constant factor and
is the inverse of the usual definition,
but this does not affect its scaling properties.
The scaling of $g_R$ is extremely simple:
\be
{g}_{R} = f ( L^{ 1/ \nu } t ) ,
\label{eq9}
\ee
where $f$ denotes a universal scaling function, $L$ is the linear
extent of the lattice and $ t =
(\lambda^{-1} -{ \lambda_c}^{-1})\lambda_c $
is the reduced coupling ("temperature").
The subscript $c$ refers to the infinite volume critical coupling.

We can determine $\lambda_c$ by noting that,
according to Eq. (\ref{eq9}), the curves of $g_R$ for different
lattice sizes should cross at $\lambda_c$, up to scaling violations
visible on too small lattices.
The simulation
results are shown in Fig. 2.
The reweighting analysis gives crossings at
$\lambda_c= 0.820(2)$ for $8^3$ and $12^3$, $\lambda_c=0.817(1)$
for $12^3$ and
$16^3$, $\lambda_c=0.815(1)$ for $16^3$ and $24^3$ and
$\lambda_c=0.817(3)$ for $24^3$ and $32^3$.
We thus conclude that the GN model has a second
order phase transition $\lambda_c = 0.815(3)$.
The renormalised coupling at $\lambda_c$ is $(g_R)_c = 0.473(4)$.

The usual way to avoid refering
to $\lambda_c$ in the critical exponents determination
is to use thermodynamic quantities
that peak in the scaling region. This is possible since, according to
the scaling ansatz, $L^{1/\nu}t$ is constant at the maxima.
Unfortunately, there are no quantities whose scaling behaviour
provide a direct estimate of the exponent $\nu$.
Therefore, if one tries to measure $\nu$ one also has
to specify the value of $t$, and thus $\lambda_c$.

We can relax this requirement by noting that the scaling
formula derived above is
independent of the critical coupling and is valid in the
whole critical region as long as the scaling violations can be
neglected. Hence, we can perform a scaling
analysis to quantities which do not
necessarily peak at the critical
coupling. Moreover, the latter need not even be specified:
invering Eq. (\ref{eq9}), $L^{1/\nu}t$ can be expressed as a
function of $g_R$ and the finite size analysis can be made at
constant value of $g_R$ \cite{Engels92b}.
We have to pay a price, though, since measuring $g_R$ can be demanding.
It is also crucial to eliminate $t$, since its
uncertainty contributes a lot to the
errors in the exponents.

In order to extract the critical exponent $\nu$, we consider
the logarithmic derivative of $g_R$
with respect to the reduced coupling
\be
D\equiv {{ \partial {\rm ~ln} ( g_R)} \over  \partial  t} =
L^{ 1/ \nu } {\frac{f '( L^{1/\nu } t )}{f( L^{1/\nu } t)}}
\equiv L^{ 1/ \nu } F( L^{1/\nu } t ),
\ee
with $F$ a new universal scaling function.
Inverting the scaling equation of $g_R$ for
$L^{1/\nu}t$ we obtain
\be
D = L^{ 1/\nu } G( g_R),
\ee
where $G$ is a scaling function.

On the other hand, $D$ is a correlator
of powers of the
$\sigma$-field and the bosonic energy $S = 1/2 \sum_n \sigma_n^2$
from the definition of average quantities :
\be
D
= \left\langle{ S }\right\rangle
+ {\frac{\left\langle{  S \sigma^4  }\right\rangle}
        {\left\langle{  \sigma^4    }\right\rangle}}
-2{\frac{\left\langle{ S \sigma^2   }\right\rangle}
        {\left\langle{ \sigma^2     }\right\rangle}},
\ee
which can thus be measured numerically (i.e. by means of MC simulations).

 From this measurement, we determine $\nu$ from a fit
at constant $g_R$ to
\be
{\rm ln} ~{D}|_{g_R={\rm const}} =
{1/\nu \ {\rm ln}~ \left[ L\right]} + {\rm const}
\ee
In Fig. 3, the value of $\nu$
is shown as a function $g_R$, together with the corresponding
$\chi^2$ value of the fit.
The scaling behaviour is realized for the entire fitting range
\be
\chi^2 < 0.5
\ee
for three degrees of freedom.
The errors on $\nu$ are coming from a fit which uses the
errors of the original data. These were obtained by a
jackknifed reweighting analysis. The estimate obtained
with $g_R = (g_R)_c = 0.473$ is
\be
\nu = 1.00(4).
\ee
Notice that the dependence on $(g_R)$ is indead weak,
and that we did not have to specify $\lambda_c$
in our fits. To see how well our data is actually scaling we
display $g_R$ versus $t L^{1/\nu}$ with our MC estimates of
$\nu$ and $\lambda_c$ in Fig. 4.

Using the hyperscaling relations, only two exponents are independent.
As a second exponent
we choose $2-\eta = \gamma/\nu$. This
governs the behaviour of the susceptibility $\chi$ near the critical
point
\be
\chi =\left\langle{{\sigma
}^{2}}\right\rangle-{\left\langle{\sigma }\right\rangle}^{2} =
L^{\gamma/\nu} g( L^{1/\nu} t),
\label{eq12}
\ee
where $g$ denotes a scaling function.
However, in numerical simulations there is a problem concerning
the susceptibility: on finite lattices
the average of the $\sigma$ field is
always zero.

The use of absolute values in definition
(\ref{eq12}) could distort the scaling behaviour and may
lead to wrong exponents. To overcome this we used the
susceptibility on the symmetric side \cite{Engels92a}
 $\lambda > \lambda_c$ where
\be
\chi =\left\langle{\sigma^2}
\right\rangle = L^{\gamma/\nu} g( L^{1/\nu} t).
\ee

Eliminating again $t$
as in the case for $D$ (and $g_R$), we obtain
\be
\ln{ \left\langle{\sigma^2}
\right\rangle}|_{g_R={\rm const}} ={\gamma\over \nu} \ln{L}+ {\rm const}.
\la{eq14}
\ee
and get from a fit to the measured $L$ dependence
\be
{\gamma\over \nu} = 1.246(8).
\la{eq15}
\ee
Fig. 5 shows the results of the fit. Notice that the fit is
valid for the whole critical range ($\chi^2 < 0.4$ for 3 d.o.f) and
the deviation as a function of $g_R$ is very small as expected.
 The value we
quote is taken at $ (g_R)_c = 0.473$.
 From Fig. 6 one can see that
the scaling of the data is excellent
with our values of exponents: within error bars all
the data from different lattice sizes lie on the same curve.

As a check of consistency with hyperscaling,
we can measure other critical exponents.
The expectation value of the sigma field acts as an order
parameter for the discrete chiral symmetry \eq{1d},
which is preserved on the lattice.
On a finite lattice, the absolute value of $\sigma$ yields
an estimate for the combination
$\beta/\nu$, which is shown in Fig. 7.
At
$g_R = (g_R)_c = 0.473$ we get
\be
{\beta\over \nu} = 0.877(4),
\la{eq15a}
\ee
with $\chi^2 < 0.3$ for 3 d.o.f.
With this value of exponents the quality of scaling is again
excellent, as one can see from Fig. 8.

The combination
$\beta/\nu$ is connected to
$\gamma/\nu$ through the hyperscaling relation
\be
{\frac{\beta}{\nu}}={\frac{1}{2}} ( d - {\frac{\gamma}{\nu}} ).
\ee
Using the value of $\gamma/\nu$ given in (\ref{eq15}) this gives
$\beta/\nu = 0.877(4)$, which is in complete
agreement with estimate obtained above (\ref{eq15a}).
This agreement is noteworthy since we used the absolute value
of $\sigma$, which can lead to a distortion
of the scaling relation. At least in our case we see that it does not.
Also, the definition of the susceptibility with
the absolute value of $\sigma$
leads to identical results.
However, for the standard method of measuring the critical
exponents from the scaling behaviour of thermodynamic
quantities at their peak values, the usage of the absolute
value $\left\langle \sigma \right\rangle$ may result
in a change in the position of the peaks
and thus make the scaling analysis dubious.

The heat capacity should give the combination $\alpha/\nu$. The
hyperscaling relation predicts a
value of $-1$. This means that heat capacity does not diverge
at the critical point. In fact, it is dominated by its regular part
which makes it impossible to extract $\alpha/\nu$.
In order to do this
we would need the second derivative of
the heat capacity with respect to $t$. This
quantity would diverge as $t^{-(\alpha/\nu+2)} \sim t^{-1}$.
Unfortunately the quality of the MC data deteriorates as higher order
derivatives of the free energy are taken: the 4th derivative is out
of reach in the present simulation.

All of the previous analysis relied
heavily on reweighting the data from a finite set
of couplings to a very dense set of couplings. This enabled us to
accurately explore the dependence of the variables on the renormalised
coupling $g_R$.
We note that both our method of analysis and the number of trajectories used
allow us to achieve a better determination of the critical exponents
than was achieved in a comparable analysis
for $N_f=12$
\cite{Hands92}.

\section{Conclusions}

We have performed a high statistics simulation of the 3d GN model with
two flavours of 4-spinors. We show that it has a second order phase
transition at $\lambda = 0.815(3)$. Hence, it leads to a continuum
field theory, which is characterised by  critical exponents
which we have measured.
This proves numerically that the
GN model is renormalisable in three dimensions,
even for a small number of flavours.
The transition point is close to the $1/N_f$ expectation $\lambda_c^1=0.80$,
but the $1/N_f^2$ correction can be as
significant as in 2d calculation \cite{belanger}.
Table~\ref{table2} displays the results from our simulations together with
estimates obtained by other methods:
the $\epsilon$-expansion is for the HY model to first order
by Zinn-Justin \cite{Justin92}, and to second order as presented above,
the $1/N_f$-expansion for the GN model  calculated
to one loop by Hands et al. \cite{Hands92} and to order $1/N^2$ by Gracey
\cite{Gracey91,Gracey92,Gracey93} and by Derkachov et al. \cite{sacha}.
The second order contributions of the HY
$\gamma / \nu$ and the GN $\nu$ are large and the
corresponding expressions have been resummed as explained in Sect. 2.4.

\begin{table}
\center
\caption{The critical exponents obtained from different methods.}
\center
The numbers with a star are obtained with resummation.
\smallskip
\label{table2}
\begin{tabular}{lrrrrr} \hline
Exponent     & \multicolumn{1}{r}{$MC$}
                        & \multicolumn{1}{r}{$\epsilon$}
                        & \multicolumn{1}{r}{$\epsilon^2$}
                        & \multicolumn{1}{r}{$1/N_f$}
                        & \multicolumn{1}{r}{$1/N_f^2$}\\
\hline
$\nu$    &$1.00(4) $&$0.9545 $&$0.9480 $&$1.135$&$0.903*$\\
$\gamma / \nu$ &$1.246(8)$&$1.4285$&$1.237*$&$1.270$&$1.2559$\\
\hline
\end{tabular}
\end{table}

The striking feature of the data is that the HY $\epsilon$-expansion results
at the two-loop level are in very good agreement with the simulation
values. Even without resummation, which can be found quite arbitrary, the
direct result of $\gamma / \nu$ is not very far from the data point as
seen in the Fig. 1.
The GN second order $1/N_f$-expansion works very well for $\gamma/\nu$
which has a small  $1/N_f^2$ correction.
Concerning the GN $\nu$, even though the
resummed value is not too far off from our numerical result,
the discrepancy does suggest that
higher order terms may be important. However, the agreement with
the HY $\nu$ shows that no new phenomenon appears at small $N$.

As a whole, our results strongly support the conjecture
that these models are equivalent even in
three dimensions, where they are not trivial, and that the properties
inferred from perturbation theory are valid at low fermion number.

\subsection*{Acknowledgments}
We thank John Gracey for discussions and informing us of his
second order results prior to
publication and Jean Zinn-Justin and Andr\'e Morel
for fruitful conversations. One of us (R.L.) thanks the
hospitality of Crete University where the $\epsilon^2$ expansion was
initiated and E. G. Floratos for
discussions and pointing out Ref. \cite{machacek}.
This project was supported by the Deutsche Forschungsgemeinschaft and
by the U.S. Dept. of Energy grant No.~DE-FG02-8SER40213. The
MC simulations were performed at the HLRZ, J\"ulich, and at Saclay.

\pagebreak

\subsection*{Figure captions}

\noindent{\bf Figure 1}. ~~ {Comparison of critical exponents
obtained with different
means as function of the effective fermion number $N$.
Solid lines for the $\epsilon^2$ Higgs-Yukawa model, dashed lines for the
$1/N^2$ Gross-Neveu model, dotted lines for HY $\gamma/\nu$ and GN
$\nu$ direct results (without resummation).
The data point at $N=48$ is for $\nu$ from  Ref. \cite{Hands92},
those for $N=8$ result from our simulation.}

\bigskip
\noindent{ \bf Figure 2}. ~~ {The renormalised
coupling $g_R$ as function of the coupling
$\lambda$ for lattice sizes $8^3$, $12^3$, $16^3$,
$24^3$ and $32^3$ in order of increasing slopes. The results of
simulations without the reweighting are shown as circles.}

\bigskip
\noindent{ \bf Figure 3}. ~~ {The critical
exponent $\nu$ as function of the  value of
$g_R$. The corresponding $\chi^2$ plot gives the quality of the fit.}

\bigskip
\noindent{ \bf Figure 4}. ~~ {The renormalised coupling $g_R$ as function of
$t L^{1/\nu}$ for different
lattice sizes ($8^3$ is labeled with plus, $12^3$ with octagons, $16^3$
with squares,
$24^3$ with circles and $32^3$ with diamonds).
The $\nu$ and $\lambda_c$ have the measured MC values. }

\bigskip
\noindent{ \bf Figure 5}. ~~ {The critical exponent $\gamma/\nu$ as
function of the critical
value of $g_R$. The $\chi^2$ gives the quality of the fit.}

\bigskip
\noindent{ \bf Figure 6}. ~~ {The combination
$\sigma^2 L^{3-\gamma/\nu}$ and $g_R$ as
a function of $t L^{1/\nu}$ for different
lattice sizes ($8^3$ is labeled with plus, $12^3$ with octagons, $16^3$
with squares, $24^3$ with circles and $32^3$ with diamonds).
The $\nu$, $\gamma/\nu$ and $\lambda_c$ have the measured MC values. }

\bigskip
\noindent{ \bf Figure 7}. ~~ {As in Fig. 5, but for the critical
exponent $\beta/\nu$.}

\bigskip
\noindent{ \bf Figure 8}. ~~ {As in Fig. 6, but for the combination
$|\sigma| L^{\beta/\nu}$.}
\end{document}